---------------------------------------------------------------------------
############################## .TEX FILE ###################################
---------------------------------------------------------------------------

\documentstyle[aps]{revtex}            
\newcommand{\R}{\, \mbox{R\hspace*{-2.8ex} I \hspace*{0.5ex}} } 
\newcommand{\Ra}{\, \mbox{R\hspace*{-2.95ex} I \hspace*{0.5ex}} } 
\newcommand{\C}{\mbox{C\hspace*{-1.1ex}\rule{0.15ex} {1.5ex} 
                        \hspace*{0.1ex}}} 
\newcommand{\D}{\mbox{D\hspace*{-1.1ex}\rule{0.15ex} {1.5ex} 
                        \hspace*{0.1ex}}} 

\newcommand{\RR}{\, \mbox{{\footnotesize R\hspace*{-3.0ex} I 
                        \hspace*{0.5ex}}} } 
\newcommand{\CC}{\mbox{{\footnotesize C\hspace*{-1.1ex}\rule{0.15ex} {1.5ex} 
                        \hspace*{0.1ex}}}} 
\newcommand{\DD}{\mbox{{\footnotesize D\hspace*{-1.2ex}\rule{0.15ex} {1.5ex} 
                        \hspace*{0.1ex}}}}

\begin{document}


\draft
\preprint{}
\title{The Dirac-Hestenes Lagrangian}
\author{Stefano De Leo$^{a,b}$,
        Zbigniew Oziewicz\thanks{Oziewicz is a member 
                                 of {\sl Sistema Nacional de 
                                 Investigadores}, M\'exico.}$^{c,d}$,
        Waldyr A.~Rodrigues, Jr.$^{b}$ and
        Jayme Vaz, Jr.$^{b}$}
\address{$^{a}$Dipartimento di Fisica, Universit\`a degli Studi Lecce,\\
               Istituto Nazionale di Fisica Nucleare, INFN, Sezione di Lecce\\
               via Arnesano, CP 193, 73100 Lecce, Italia\\
               {\tt deleos@le.infn.it}\\    
         $^{b}$Instituto de Matem\'atica, Estat\'{\i}stica e 
               Computa\c{c}\~ao Cient\'{\i}fica, IMECC-UNICAMP\\
               CP 6065, 13081-970  Campinas, S.P., Brasil\\
               {\tt deleo/walrod/vaz@ime.unicamp.br}\\ 
         $^{c}$Universidad Nacional Aut\'onoma de M\'exico,
               Facultad de Estudios Superiores,\\
               CP 54700 Cuautil\'an Izcalli, Apartado Postal 25, M\'exico\\
               {\tt oziewicz@servidor.unam.mx}\\    
         $^{d}$Uniwersytet Wroc{\l}awski, Instytut Fizyki Teoretycznej,\\ 
               plac Maksa Borna 9, 50204 Wroc{\l}aw, Poland\\ 
               {\tt oziewicz@ift.uni.wroc.pl} }
\date{\today}
\maketitle


\begin{abstract}

We discuss the variational principle within Quantum Mechanics in terms of 
the noncommutative even Space Time sub-Algebra,  
the Clifford $\Ra$-algebra $Cl_{1,3}^+$. 
A fundamental ingredient, in our multivectorial algebraic 
formulation, is the adoption of a $\D $-complex geometry, 
$\D \equiv \mbox{span}_{\RR} \{ 1,\gamma_{21} \}$,  
$\gamma_{21} \in Cl_{1,3}^+$.  
We derive the Lagrangian for the Dirac-Hestenes equation and  show that 
such Lagrangian must be mapped on 
$\D \otimes {\cal F}$,
where $\cal F$ denotes an $\Ra$-algebra of functions. 


{\bf 1996 PACS numbers:} 02.10.Tq, 02.10.Rn, 03.65.Pm, 11.10.Ef, 12.15.-y.


\end{abstract}


\section{Introduction}

This introduction contains a brief summary of the translation between the 
Dirac~\cite{DIR,ZB} and Dirac-Hestenes~\cite{HESP} equation. Throughout the 
paper we use the notation:
\begin{center}
\begin{tabular}{cclrcl}
$\R$                & ~is~     & the real field , &
$\otimes$           & ~$\equiv$~ & $\otimes_{\RR}~,$\\
$\C$                & ~is~     & the complex field 
$\C \equiv \mbox{span}_{\RR} \{ 1, i \}$, $i \in \C$, 
                                 $i^2=-1$ ,~~~~~~~ &
$x_{\mu}$           & $ \in$   &  ${\cal F}~,$\\
$\D$                & ~is~     & the field 
$\D \equiv \mbox{span}_{\RR} \{ 1, \gamma_{21} \}$, 
$\gamma_{21} \in Cl_{1,3}^+$, 
$\gamma_{21}^2=-1$ , ~~~~~~~ &
$\partial^{\mu}$    & $ \in$   &  $\mbox{der} {\cal F}~,$\\
${\cal F}$          & ~is~     & an $\R$-algebra of functions 
from ${\R}^4$ to $\R ~,$ &
$\partial^{\mu} x_{\nu}$ & = & $\delta^{\mu}_{\nu} \in {\cal F}~,
   ~~~~~\mbox{\footnotesize $\mu,\nu=0,1,2,3$}~.$ 
\end{tabular}
\end{center}
Let $g_{\mu \nu} = \mbox{diag} \, \left( \, +, \, -,  \, -, \, - \, \right)$ 
be the Minkowski metric, 
${\cal I} \triangleleft \left( \C \otimes Cl_{1,3} \right)$
be one-sided ideal, 
${\cal I} \approx {\C}^4$,  and $\Psi_D \in {\cal I} \otimes {\cal F}$,
\begin{equation}
\label{spi}
   \Psi_D  \equiv  
   \left( \begin{array}{c} a_1 + i b_1  \\ 
                           a_2 + i b_2  \\ 
                           a_3 + i b_3  \\ 
                           a_4 + i b_4  \end{array} 
   \right) \equiv 
   \left( \begin{array}{c} \psi_{D,1} \\ \psi_{D,2} \\ 
                           \psi_{D,3} \\ \psi_{D,4} \end{array} 
   \right)
~~~~~~~a_{m},~b_{m}  \in {\cal F}~,~~ \psi_{D,m} \in \C \otimes {\cal F}~,
~~~~~\mbox{\footnotesize $m=1,2,3,4$}~.
\end{equation}
Let $\gamma_{\mu} \in \mbox{End} ({\cal I}) \approx \mbox{Mat}_4( \C )$ be 
$4 \times 4$ complex matrices which satisfy the Dirac algebra:
\[ \gamma_{\mu}  \gamma_{\nu} + \gamma_{\nu} \gamma_{\mu} = 2 g_{\mu \nu}
   {\openone}_4~~~\in \mbox{Mat}_4( \C )~,
~~~~~\mbox{\footnotesize $\mu,\nu=0,1,2,3$}~.\]
If  $m \in {\R}^+$ is the mass of the particle, then the 
Dirac equation for a free particle $\Psi_D$ reads   
\begin{equation}
\label{dir}
i \gamma_{\mu} \partial^{\mu} \Psi_D = m \Psi_D  ~~~ 
\in {\cal I} \otimes {\cal F}~.
\end{equation}
For the  Dirac matrices, a possible choice, useful for the discussion
presented in Section~\ref{s2}, is   
\[ \gamma^{\mu} \equiv \left\{ \gamma_0 \, , \,  \gamma_k \right\}~~~~~~~
\mbox{\footnotesize $k=1,2,3$}~, \]
\begin{equation}
\label{sr}
\gamma_{0} \equiv \left( \begin{array}{cccc} 
                          1 & 0 & 0     & 0 \\
                          0 & 1 & 0     & 0 \\
                          0 & 0 & $-$ 1 & 0 \\
                          0 & 0 & 0     & $-$ 1 \end{array} \right)~,~~~
\gamma_1 \equiv \left(  \begin{array}{cccc} 
                          0 & 0 & 0     & i \\
                          0 & 0 & i     & 0 \\
                          0 & i & 0     & 0 \\
                          i & 0 & 0     & 0 \end{array} \right)~,~~~
\gamma_2 \equiv \left(  \begin{array}{cccc} 
                          0 & 0      & 0   &  $-$ 1 \\
                          0 & 0      & 1   & 0 \\
                          0 &  $-$ 1 & 0   & 0 \\
                          1 & 0      & 0   & 0 \end{array} \right)~,~~~
\gamma_3 \equiv \left(  \begin{array}{cccc} 
                          0 & 0 &  $-$ i  & 0 \\
                          0 & 0 & 0       & i \\
                      $-$ i & 0 & 0       & 0 \\
                          0 & i & 0       & 0 \end{array} \right)~.
\end{equation}

A renewed interest exists in the formulation of the Dirac theory 
in terms of the Clifford algebra~\cite{SS,CA1,CA2,CA3,CA4,CA5}. An 
interesting result is 
the possibility to write the Dirac equation in the even Space Time 
sub-Algebra, $Cl_{1,3}^+$. This result is achieved by working only 
with the general properties of the Clifford $\R$-algebra, in special the 
concept of even sub-algebra~\cite{ZEN}.

An alternative possibility in rewriting the 
Dirac equation in $Cl_{1,3}^+$, is represented by
the direct {\em translation} of the elements which characterized the standard 
``complex'' formulation. In the present Section, 
we summarize this translation and introduce the 
notion of $\D $-complex geometry~\cite{REM,HB,DR,DRV}. 

Consider a sub-algebra isomorphic to the complex field $\C$
\[ 
Cl_{1,3}^+ > 
\D \equiv  \mbox{span}_{\RR} \{ 1,\gamma_{21} \} 
\approx \C =  \mbox{span}_{\RR} \{ 1,i \} ~,~~~~~ 
\gamma_{21} \leftrightarrow i~.  
\] 
By $\D $-complex geometry we mean an $\R$-linear mapping
\begin{equation}
\label{chi}
   \chi ~:~~~ Cl_{1,3}^+ \otimes {\cal F} ~ \rightarrow ~
   \D \otimes {\cal F} ~,~~~~~ \D < Cl_{1,3}^+~, 
\end{equation}
\[ \chi \in \mbox{lin}_{\RR} \left( Cl_{1,3}^+ \otimes {\cal F} \, , \, 
                                    \D \otimes {\cal F} \right)~,~~~
   \Psi \in  Cl_{1,3}^+ \otimes {\cal F}~,~~~
   \chi \left( \Psi \right) \equiv \left( \Psi \right)_{\cal F} -
   \gamma_{21} \left( \gamma_{21} \Psi \right)_{\cal F}~,
\]
where the subscript $_{{\cal F}}$ denotes the mapping on 
$\R \otimes {\cal F} \approx {\cal F}$ of the quantity within the 
brackets.

The Dirac spinor fields are elements of the $\C $-space 
${\C}^4 \otimes {\cal F}$, 
thus are characterized by 8 real functions of four real variables, 
e.g.~\cite{MAG},   
\[ \mbox{dim}_{\RR} {\cal I} = \mbox{dim}_{\RR} Cl_{1,3}^+ = 8~.\] 
The possible basis of the Clifford $\R $-algebra $Cl_{1,3}^+$ is
\begin{center}
\begin{tabular}{cc}
$1$~, & \\
$\gamma_{01}$,~$\gamma_{02}$,~$\gamma_{03}$~, & \\
$\gamma_{21}$,~$\gamma_{31}$,~$\gamma_{23}$~, & \\
$\gamma_{5} \equiv \gamma_{0123} $ & ~~~~$\in Cl_{1,3}^+~.$~~~~
\end{tabular}
\end{center}
An arbitrary element in $Cl_{1,3}^+ \otimes {\cal F}$ can be written  
as
\begin{equation}
\label{e1}              \alpha_0 + \gamma_{01} \alpha_1  +
                  \gamma_{02} \alpha_2  + \gamma_{03} \alpha_3  +
                  \gamma_{21}  \alpha_4  + \gamma_{31} \alpha_5  +
                  \gamma_{23} \alpha_6  + \gamma_{5} \alpha_7 ~,  
~~~~~~~~~~\alpha_m \in {\cal F}~,
~~~~~\mbox{\footnotesize $m=0,...,7$}~.
\end{equation}
The Hestenes spinor, solution of the Dirac-Hestenes equation, 
\[
\psi_{H,m} = a_m  + \gamma_{21} b_m \in \D \otimes {\cal F}~,
~~~~~\mbox{\footnotesize $m=1,2,3,4$}~, 
\]
\begin{equation}   
\label{e3}
      \Psi_H \equiv   \psi_{H,1}  + \gamma_{31} \psi_{H,2} + 
            \gamma_{5} \left( \psi_{H,3} + \gamma_{31} \psi_{H,4} \right)~
\in Cl_{1,3}^+ \otimes {\cal F}~,
\end{equation} 
will represent the counterpart in the even Space Time sub-Algebra of 
the complex Dirac spinor  $\Psi_D$. The following {\em isomorphism}, 
\begin{equation}
\label{rho}
\rho~:~~~{\cal I} \otimes {\cal F} 
~ \rightarrow ~ Cl_{1,3}^+ \otimes {\cal F}~,
\end{equation}
requires the identification 
\[ \rho \in \mbox{lin}_{\RR} \left( {\cal I} \otimes {\cal F} \, , \, 
                                    Cl_{1,3}^+ \otimes {\cal F} \right)~,~~~
i \in {\cal I}~,~~~ \gamma_{21} \in Cl_{1,3}^+~,~~~\rho(i)=\gamma_{21}~. 
\]

Let $\beta$ be the main anti-automorphism of the Clifford $\C $-algebra 
$\C \otimes Cl_{1,3}$, then there
exists hermitian sesquilinar form~\cite{CRU} 
in the space of the Dirac spinors
\begin{equation}
\label{1}
   h \, : ~ {\cal I}^+ \otimes {\cal I} ~ \rightarrow ~ \C ~, ~~~~~~~ 
   h \in {\cal I}^* \otimes \left( {\cal I}^+ \right)^*~, ~~~~~~~
   h \left( {\cal I}^+ \otimes {\cal I} \right) \equiv 
     \left( \beta {\cal I} \right) {\cal I} \in \C~.
\end{equation}
Exists a basis in ${\cal I}$ such that 
\begin{eqnarray} 
\label{2}
  \Phi^{\dag}_D h \Psi_D & \equiv & 
   \left( \varphi_{D,1}^* , \, \varphi_{D,2}^* , \, \varphi_{D,3}^* , 
          \, \varphi_{D,4}^*  
   \right)
   \left( \begin{array}{cccc} 1 & 0 & 0    & 0\\
                              0 & 1 & 0    & 0\\
                              0 & 0 & $-$1 & 0\\
                              0 & 0 & 0    & $-$1
          \end{array} \right) 
   \left( \begin{array}{c} \psi_{D,1} \\ \psi_{D,2} \\ \psi_{D,3} \\ 
                           \psi_{D,4} \end{array} 
   \right) \nonumber \\
  & \equiv  & 
   \varphi_{D,1}^* \psi_{D,1} + \varphi_{D,2}^*  \psi_{D,2} - 
   \varphi_{D,3}^* \psi_{D,3} - \varphi_{D,4}^*  \psi_{D,4}~, 
\end{eqnarray}
\[ \varphi_{D,m} \, , \, \psi_{D,m}  \in \C \otimes {\cal F}
~,~~~\mbox{\footnotesize $m=1,2,3,4$}~.
\] 
We recall that 
$\gamma_{\mu} \in \mbox{End} ({\cal I}) \approx \mbox{Mat}_4( \C )$,
\[ \gamma_{\mu} \, : ~ {\cal I} ~ \rightarrow ~ {\cal I} ~, ~~~~~~~ 
   \gamma_{\mu} \in {\cal I} \otimes {\cal I}^*~.\]
Under change of basis
\[
\gamma_{\mu} ~\rightarrow ~  {\cal S} \gamma_{\mu} {\cal S}^{-1}~,~~~~~
h            ~\rightarrow ~  {\cal S}^t h {\cal S}^*~.
\]
Therefore, $h$ and $\gamma_{\mu}$ are different tensors and the identification 
$h=\gamma_0$ is not correct.

In translating the previous hermitian product~(\ref{1},\ref{2}) 
in the  $Cl_{1,3}^+$ formalism, we
need to single out the conjugation which characterizes the standard 
hermitian conjugate and impose an appropriate geometry.  
In order to translate $\Phi_D^{\dag}$  into $Cl_{1,3}^+ \otimes {\cal F}$, 
we must determine the possible automorphisms ($\alpha$) 
and anti-automorphisms ($\beta$) of $Cl_{1,3}^+$:
\[ \alpha \in \mbox{aut} \left( Cl_{1,3}^+ \otimes {\cal F} \right)~,~~~
   \beta \in \mbox{anti-aut} \left( Cl_{1,3}^+  \otimes {\cal F} \right) ~,
\]
\[ \alpha \left(  \Psi_H \Phi_H  \right) = 
   \alpha \left(  \Psi_H \right) \alpha \left( \Phi_H  \right)~,~~~
   \beta  \left(  \Psi_H \Phi_H  \right) = 
   \beta  \left(  \Phi_H \right) \beta \left( \Psi_H  \right)~.
\]
We find,  
\begin{eqnarray*}
\alpha ~:~~~& \mbox{grade involution ,} & 
~~~
\gamma_{0i} \rightarrow - \gamma_{0i}~,~~
\gamma_{ij} \rightarrow + \gamma_{ij}~,~~ 
\gamma_5    \rightarrow - \gamma_5~,\\
               &        & ~~~~~\alpha \left( \Phi_H \right) \equiv 
                          \widehat{\Phi}_H =  
                         \varphi_{H,1} + \gamma_{31} \varphi_{H,2} - 
       \gamma_{5} \left( \varphi_{H,3} + \gamma_{31} \varphi_{H,4} 
                  \right)~,\\
\beta ~:~~~ & \mbox{reversion ,} & 
~~~
\gamma_{0i} \rightarrow + \gamma_{0i}~,~~
\gamma_{ij} \rightarrow - \gamma_{ij}~,~~ 
\gamma_5    \rightarrow - \gamma_5~,\\
                &       & ~~~~~\beta \left( \Phi_H \right) \equiv 
                          \widetilde{\Phi}_H  =  
                         \varphi_{H,1}^*   - \varphi_{H,2}^*  \gamma_{31} -
       \gamma_{5} \left( \varphi_{H,3}^*   -  \varphi_{H,4}^* \gamma_{31} 
                  \right)~,\\
\alpha \circ \beta ~:~~~ & \mbox{Clifford conjugation ,} & 
~~~
\gamma_{0i} \rightarrow - \gamma_{0i}~,~~
\gamma_{ij} \rightarrow - \gamma_{ij}~,~~ 
\gamma_5    \rightarrow + \gamma_5~,\\
               &        & \alpha \circ \beta \left( \Phi_H \right) \equiv  
                         \overline{\Phi}_H  =  
                         \varphi_{H,1}^*  - \varphi_{H,2}^*  \gamma_{31} + 
       \gamma_{5} \left( \varphi_{H,3}^*  -  \varphi_{H,4}^*  \gamma_{31} 
                  \right)~,
\end{eqnarray*}
\[ 
\mbox{{\footnotesize $i,j=1,2,3$~,~~$i\neq j$}}~,~~~
\varphi_{H,m}  \in \D \otimes {\cal F}
~,~~~~~\mbox{\footnotesize $m=1,2,3,4$}~.
\] 
The hermitian sesquilinear form  
$\Phi_D^{\dag} h  \Psi_D \in \C \otimes {\cal F}$ 
can be translated by using the 
reversion and grade involution and adopting a 
$\D $-complex mapping~(\ref{chi})  
\begin{equation}
\label{l1}
\C \otimes {\cal F} \ni ~
\Phi_D^{\dag} h \Psi_D ~~~\leftrightarrow ~~~
\chi \left(  \widetilde{\Phi}_H \widehat{\Psi}_H  \right) \equiv     
\left[ \widetilde{\Phi}_H \widehat{\Psi}_H \right]_{{\cal F}} - 
\gamma_{21} \left[ \gamma_{21} \widetilde{\Phi}_H  
\widehat{\Psi}_H \right]_{{\cal F}}
~\in \D \otimes {\cal F}~.
\end{equation}

The $\D $-complex geometry, mapping on 
$\D \otimes {\cal F}$,  
is also justified by the following 
argument: We can define an anti-self-adjoint operator, $\vec{\partial}$, 
with all the properties of a translation operator, but, by imposing 
non-complex geometries, there is no corresponding self-adjoint operator 
with all the properties expected for a momentum operator~\cite{ADL}. 
The identification of $i$ with the bivector $\gamma_{21}$ 
gives us two possibilities in defining the momentum operator, 
respectively left and right action of the bivector $\gamma_{21}$, 
\[ \rho \left( i \Psi_D \right) =  \gamma_{21} \Psi_H
~~~~~\mbox{or}~~~~~
\rho \left( i \Psi_D \right) =  \Psi_H \gamma_{21} ~,~~~~~
[ \gamma_{21} \, , \, \Psi_H ] \neq 0~, 
\]
and thus we can define the following momentum operators 
\[ $-$\gamma_{21} \vec{\partial} \Psi_H ~~~~~\mbox{or}~~~~~
   $-$\vec{\partial} \Psi_H   \gamma_{21}~. \] 
By  introducing the concept of left/right operators
\[ 
{\cal O}^{l,r} \in \mbox{End} \left( Cl_{1,3}^+ \otimes {\cal F} \right)~,~~~
   {\cal O}^l \Psi_H  \equiv {\cal O} \Psi_H ~,~~~
   {\cal O}^r \Psi_H  \equiv \Psi_H {\cal O}~,~~~~~~~
   \Psi_H \in Cl_{1,3}^+ \otimes {\cal F}~,
\]
necessary within noncommutative algebraic structures, where we must 
distinguish between the left and right multiplication, we can express the 
momentum operator in $Cl_{1,3}^+ \otimes {\cal F}$ as 
\[ $-$\gamma_{21}^l \vec{\partial} \, \equiv \,
   $-$\gamma_{21}^l \otimes \vec{\partial} ~~~~~\mbox{or}~~~~~
   $-$\gamma_{21}^r \vec{\partial} \, \equiv \, 
   $-$\gamma_{21}^r \otimes \vec{\partial}~.\]

In translating the Dirac equation, the first choice 
($\gamma_{21}$-left action) must be rejected because such an 
operator, due to the term 
\[ \gamma_{01}^l \partial_{x_1} + 
   \gamma_{02}^l \partial_{x_2} + 
   \gamma_{03}^l \partial_{x_3} 
~~
\in \mbox{End} \left( Cl_{1,3}^+ \otimes {\cal F} \right)~, \]
does not commute with the Dirac-Hestenes Hamiltonian, ${\cal H}_H$.  

In the second case ($\gamma_{21}$-right action), 
the operator $\gamma_{21}^r \vec{\partial}$ is real on the left,
thus commutes with ${\cal H}_H$,
\[ \left[ \gamma_{21}^r \vec{\partial} \, , 
    \, {\cal H}_H \right] \, \Psi_H   \equiv
     \vec{\partial} \left( {\cal H}_H \Psi_H \right) \gamma_{21} -
     {\cal H}_H \vec{\partial} \Psi_H  \gamma_{21} = 0~.
\]
It remains to prove the hermiticity of the momentum operator,  
\( $-$ \gamma_{21}^r \vec{\partial} \). To do it, we need to define an
appropriate mapping for scalar products. If probability 
amplitudes are assumed to be element of non division algebras (in this case
$Cl_{1,3}^+$), we cannot give a satisfactory probability 
interpretation~\cite{ADL}.

It is seen that a $\D $-complex mapping~(\ref{chi}) 
\[ 
\chi \left( \langle \Phi_H \mid \Psi_H \rangle \right) \equiv 
\left( \int d^3 x ~\tilde{\Phi}_H \Psi_H  \right)_{\DD \otimes {\cal F}}~,
\] 
overcomes the previous problem and gives the required hermiticity properties
for the momentum operator
\begin{equation}
\label{hp}
\chi \left( \langle \Phi_H \mid \vec{\partial} 
\Psi_H  \gamma_{21} \rangle \right) = 
\chi \left( \langle \vec{\partial} 
\Phi_H \gamma_{21} \mid \Psi_H 
\rangle \right) ~. 
\end{equation} 
Eq.~(\ref{hp}) implies
\[
\left( \int d^3 x  ~\tilde{\Phi}_H  \vec{\partial} \Psi_H  
\right)_{\DD \otimes {\cal F}} 
\gamma_{21} = - \gamma_{21} 
\left( \int d^3 x 
~\vec{\partial} \tilde{\Phi}_H  \Psi_H  \right)_{\DD \otimes {\cal F}}~.
\]
Now, to prove the hermiticity of our momentum operator it is sufficient to
perform integration by parts and use the $\D $-complex mapping. 

We conclude this section by observing that there is a {\em difference} in 
translating complex operators and states. For example, the complex imaginary
unit $i$ can be interpreted as operator
\[ i \openone_4 \in \mbox{End} \left( \C^4 \right)~, \]
or state
\[ 
\left( \begin{array}{c} i\\ 0 \\ 0\\ 0 \end{array} \right)~,~~~
\left( \begin{array}{c} 0\\ i \\ 0\\ 0 \end{array} \right)~,~~~
\left( \begin{array}{c} 0\\ 0 \\ i\\ 0 \end{array} \right)~,~~~
\left( \begin{array}{c} 0\\ 0 \\ 0\\ i \end{array} \right)~~~~\in {\cal I}~.
\]
The translation will be respectively,
\[ \gamma_{21}^r \in \mbox{End} \left( Cl_{1,3}^+ \right)~,  \]
or
\[ 
\gamma_{21}~,~~~\gamma_{31} \gamma_{21} = \gamma_{32}~,~~~
\gamma_5 \gamma_{21} = \gamma_{03}~,~~~
\gamma_5 \gamma_{31} \gamma_{21} = \gamma_{01}~~~~\in Cl_{1,3}^+~.
\]
Let $\rho^{End}$ be the endomorphism 
linear mapping
\begin{equation}
\label{rhoe}
\rho^{End}~:~~~\mbox{End} \left( {\cal I} \right)  
~ \rightarrow ~ \mbox{End} \left( Cl_{1,3}^+ \right)~.
\end{equation}
We require
\[ \rho ( i \Psi_D ) = \rho ( \Psi_D i )~,~~~~~ [ i \, , \, \Psi_D ] = 0~. 
\]
The previous relation is satisfied because
\begin{eqnarray*}
\rho ( i \Psi_D ) & = & \rho^{End} ( i \openone_4 ) \rho ( \Psi_D ) = 
                        \gamma_{21}^r \Psi_H = \Psi_H \gamma_{21}~,\\
\rho ( \Psi_D i ) & = & \rho ( \Psi'_D ) = \Psi'_H = \Psi_H \gamma_{21}~.
\end{eqnarray*}

\section{Dirac Equation}
\label{s2}

Once obtained the translation from the Dirac spinor field 
$\Psi_D \in {\cal I} \otimes {\cal F}$ to the Hestenes spinor field  
$\Psi_H \in Cl_{1,3}^+ \otimes {\cal F}$, 
it is possible to translate the standard complex Dirac equation 
into the even Space Time sub-Algebra.

For convenience, we multiply the left and right hand of Eq.~(\ref{dir})
by $\gamma_0$,
\begin{equation}
\label{dir2}
i \left( \partial_t + \gamma_0 \gamma_k \partial_k \right) \Psi_D
= m \gamma_0 \Psi_D~~~~~~~\mbox{\footnotesize $k=1,2,3$}~.
\end{equation}
We shall prove that this equation can be translated in the 
$Cl_{1,3}^+$ formalism, $\rho$ given by (\ref{rho}),
\begin{equation}
\label{map}
\rho \left[ i \left( \partial_t + \gamma_0 \gamma_k \partial_k \right) 
\Psi_D \right] 
= m \rho \left( \gamma_0 \Psi_D \right)~~~~~~~\mbox{\footnotesize $k=1,2,3$}~.
\end{equation}
In the previous Section, we have established the following maps
\begin{equation}    
\label{is1}
\rho \left( \Psi_D \right) = \Psi_H
~~~\mbox{and} ~~~ 
\rho \left( i \Psi_D \right) = \gamma_{21}^r \Psi_H \equiv 
\Psi_H \gamma_{21}~,
\end{equation}
thus to complete the translation of Eq.~(\ref{dir2}) in the $Cl_{1,3}^+$ 
formalism, it remains to calculate 
\[    \rho \left( \gamma_0 \Psi_D \right)
~~~\mbox{and} ~~~ 
\rho \left( \gamma_0 \gamma_k \Psi_D \right) 
~~~~\mbox{\footnotesize $k=1,2,3$}~.
\]
By using the explicit form of the Dirac matrices, given in 
Eq.~(\ref{sr}), we find 
\[
\gamma_0 \gamma_1 \Psi_D \equiv i \, 
\left( \begin{array}{c} \psi_{D,4} \\ \psi_{D,3} \\ 
                    $-$ \psi_{D,2} \\ $-$ \psi_{D,1} 
       \end{array} \right)~,~~~
\gamma_0 \gamma_2 \Psi_D \equiv  
\left( \begin{array}{c} $-$ \psi_{D,4} \\ \psi_{D,3} \\ 
                    \psi_{D,2} \\ $-$ \psi_{D,1} 
       \end{array} \right)~,~~~
\gamma_0 \gamma_3 \Psi_D \equiv i \, 
\left( \begin{array}{c} $-$ \psi_{D,3} \\ \psi_{D,4} \\
                    \psi_{D,1} \\ $-$ \psi_{D,2} 
       \end{array} \right)~,~~~
\gamma_0 \Psi_D \equiv  
\left( \begin{array}{c} \psi_{D,1} \\ \psi_{D,2} \\ 
                    $-$ \psi_{D,3} \\ $-$ \psi_{D,4} 
       \end{array} \right)~.
\]
The task is to obtain their counterpart 
in $Cl_{1,3}^+ \otimes {\cal F}$. The solution is
\begin{center}
\begin{tabular}{rcrcl}
$\gamma_{01} \Psi_H$ & $~=~$ & 
$\gamma_{21} \gamma_5 \gamma_{31} \Psi_H $ & $~\equiv ~$ &
$ \left[ \psi_{H,4} + \gamma_{31} \psi_{H,3} 
         - \gamma_5 \left( \psi_{H,2} + \gamma_{31} \psi_{H,1} \right) 
  \right] \, \gamma_{21}$~,\\
$\gamma_{02} \Psi_H$ & $~=~$ & 
$ - \gamma_5 \gamma_{31} \Psi_H $ & $~\equiv ~$ &
$ - \psi_{H,4} + \gamma_{31} \psi_{H,3} 
         + \gamma_5 \left( \psi_{H,2} - \gamma_{31} \psi_{H,1} \right) $~,\\
$\gamma_{03} \Psi_H$ & $~=~$ & 
$\gamma_{21} \gamma_5 \Psi_H $ & $~\equiv ~$ &
$ \left[ - \psi_{H,3} + \gamma_{31} \psi_{H,4} 
         + \gamma_5 \left( \psi_{H,1} - \gamma_{31} \psi_{H,1} \right) 
  \right] \, \gamma_{21}$~,\\
$\alpha ( \Psi_H )$ & $~=~$  & $\widehat{\Psi}_H$ & $~\equiv ~$ &
$ \psi_{H,1} + \gamma_{31} \psi_{H,2} 
         - \gamma_5 \left( \psi_{H,3} + \gamma_{31} \psi_{H,4} \right)$~.
\end{tabular} 
\end{center}
We have now all the needed tools to complete the translation of the 
Dirac equation in the $Cl_{1,3}^+$ formalism. The isomorphisms
\[ 
\rho \left( \gamma_0 \Psi_D \right) = \alpha (\Psi_H ) = \widehat{\Psi}_H
~~~\mbox{and} ~~~ 
\rho \left( \gamma_0 \gamma_k \Psi_D \right) = \gamma_{0k} \Psi_H   
~~~~\mbox{\footnotesize $k=1,2,3$}~,
\]
together with Eq.~(\ref{is1}), allow to write the
$Cl_{1,3}^+$ counterpart of Eq.~(\ref{dir2}). 
Finally, the translated Dirac-Hestenes equation reads 
\begin{equation}
\label{he}
\left( \partial_t + \gamma_{0 k} \partial_k \right) \Psi_H \gamma_{21} 
= m \widehat{\Psi}_H~~~\in Cl_{1,3}^+ \otimes {\cal F}~,
~~~~~~~\mbox{\footnotesize $k=1,2,3$}~.
\end{equation}

The choice of the Dirac matrices~(\ref{sr}) 
was {\em ad hoc} to obtain a simple 
translation for the complex Dirac matrices $\gamma_0 \gamma_k$
\[
\rho \left( \gamma_0 \gamma_k \Psi_D \right) = \gamma_{0k} \Psi_H   
~~~~\mbox{\footnotesize $k=1,2,3$}~.
\]
What happens if we change our basis, 
$\gamma_{\mu}^{new} = {\cal S} \gamma_{\mu} {\cal S}^{-1}$? We shall show 
that it is possible to construct a set of translation rules which enables
us to obtain for a generic $4\times 4$ complex matrix its counterpart in the 
even Space Time sub-Algebra. 
Thus, the problem concerning the translation of $\gamma_{\mu}^{new}$ is
overcome.

A generic $4 \times 4$ complex matrix is characterized by 32 real elements, 
whereas $\mbox{dim}_{\RR} Cl_{1,3}^+ = 8$, thus it seems that we have not the 
needed real freedom degrees to perform our translation. Nevertheless, we must
observe that the space of the Hestenes spinors
\[
\Psi_H \in Cl_{1,3}^+  \otimes {\cal F}~,~~~~~
Cl_{1,3}^+  \otimes {\cal F}~~\mbox{is}~~Cl_{1,3}^+ \mbox{-bimodule}~.
\]
This implies, due to the noncommutativity of the Clifford algebra
$Cl_{1,3}^+$, a left/right action on $\Psi_H$. So,  we must 
consider with the standard 8 left generators
\begin{equation} 
\label{lg}
  1^l ~ , ~ \gamma_{01}^l ~ , ~ \gamma_{02}^l ~ , ~ \gamma_{03}^l ~ ,
        ~ \gamma_{21}^l ~ , ~ \gamma_{31}^l ~ , ~ \gamma_{23}^l ~ ,
        ~ \gamma_{5}^l ~~~\in \mbox{End} \left( Cl_{1,3}^+ \right)~, 
\end{equation}
the right generators
\begin{equation}
\label{rg}
\gamma_{21}^r  ~,~ \gamma_{31}^r ~ ,~ \gamma_{23}^r 
~~~\in \mbox{End} \left( Cl_{1,3}^+ \right)~.
\end{equation}
It is not necessary to consider 
$\gamma_{01}^r$, $\gamma_{02}^r$, $\gamma_{03}^r$, because these operators 
can be obtained from the previous ones~(\ref{rg}) by $\gamma_5$
multiplication, $\left[ \gamma_5 \, ,\, Cl_{1,3}^+ \right] = 0$. By using
left~(\ref{lg}) and right~(\ref{rg}) generators we can write the following
operators
\[ o_1^l + o_2^l \gamma_{21}^r + o_3^l \gamma_{31}^r + o_4^l \gamma_{23}^r~, 
   ~~~~~ o_m^l \in \mbox{End} \left( Cl_{1,3}^+ \right)~,
   ~~~~~ \mbox{\footnotesize $m=1,2,3,4$}~,
\]
characterized by 32 real parameters. {\it This does not imply necessarily the
possibility of a translation}. In the standard Dirac theory, the operators 
are given in terms  of $4 \times 4$ complex matrices and so represent 
$i$-complex linear operators
\[
{\cal O}_D \left[ \Psi_D (a+ib) \right] = 
\left( {\cal O}_D \Psi_D \right) (a+ib)~,~~~~~a,b \in \R~.
\]
To perform our translation we must require a $\D $-complex 
linearity for our operators
\[
{\cal O}_H \left[ \Psi_H (a+ib) \right] = 
\left( {\cal O}_H \Psi_H \right) (a+ \gamma_{21} b)~.
\]
This implies that the only acceptable right generator is $\gamma_{21}^r$.
the problem is now the lack of 16 real freedom degrees
\[ 
\mbox{dim}_{\RR} \left( o_1^l + o_2^l \gamma_{21}^r \right) = 16~.
\]
The solution is achieved by recalling that in the Clifford algebra 
$Cl_{1,3}^+$, the grade involution 
$\alpha \in \mbox{aut} \left( Cl_{1,3}^+ \otimes {\cal F} \right)$ 
represents a $\D $-complex linear operator
\begin{eqnarray*}
\alpha \left[ \Psi_H (a+ \gamma_{21} ) \right] & =  & 
\alpha \left( \Psi_H \right) \alpha (a+ \gamma_{21} b)~,~~~~~~~
\alpha (\gamma_{21} ) = \gamma_{21} ~,~~~~~a,b \in \R~,\\
 & = & \alpha \left( \Psi_H \right) (a+ \gamma_{21} b)~.
\end{eqnarray*}
To obtain the set of translation rules it is sufficient to give explicitly 
the matrix counterpart of the operators
\begin{equation}
\label{mo}
1^l ~,~ \gamma_{21}^l ~,~ \gamma_{31}^l ~,~ \gamma_5^l ~,~ 
\gamma_{21}^r ~,~ \alpha
~~~\in \mbox{End}_{\DD} \left( Cl_{1,3}^+ \right)~,
\end{equation}
the others operators will be soon achieved by suitable multiplications of the
previous ones. It is evident that
\begin{equation}
\label{mo1}
1^l \leftrightarrow  \openone_4~~~~~\mbox{and}~~~~~
\gamma_{21}^r  \leftrightarrow  i \openone_4~.
\end{equation}
A computation shows that
\begin{equation}
\label{mo2}
\gamma_{21}^l \leftrightarrow 
\left( \begin{array}{cccc} i & 0     & 0 & 0\\
                           0 & $-$ i & 0 & 0\\
                           0 & 0 & i & 0\\
                           0 & 0 & 0 & $-$ i
\end{array}
\right)~,~~~
\gamma_{31}^l \leftrightarrow 
\left( \begin{array}{cccc} 0 & $-$1     & 0 & 0\\
                           1 & 0 & 0 & 0\\
                           0 & 0 & 0 & $-$1\\
                           0 & 0 & 1 & 0
\end{array}
\right)~,~~~
\gamma_{5}^l \leftrightarrow 
\left( \begin{array}{cccc} 0 & 0     & $-$1 & 0\\
                           0 & 0 & 0 & $-$1\\
                           1 & 0 & 0 & 0\\
                           0 & 1 & 0 & 0
\end{array}
\right)~,~~~
\alpha \leftrightarrow 
\left( \begin{array}{cccc} 1 & 0     & 0 & 0\\
                           0 & 1 & 0 & 0\\
                           0 & 0 & $-$ 1 & 0\\
                           0 & 0 & 0 & $-$ 1
\end{array}
\right)~.
\end{equation}                               
By using Eqs.(\ref{mo1}-\ref{mo2}) we can write the matrix counterpart for a
generic left/right generator. For example,
\begin{center}
\begin{tabular}{lcrcc}
$\gamma_{01}^l = \gamma_{21}^l \gamma_{31}^l \gamma_5^l$ &
~~~$\leftrightarrow$ ~~~ &
$ \left( \begin{array}{cccc} i & 0     & 0 & 0\\
                           0 & $-$ i & 0 & 0\\
                           0 & 0 & i & 0\\
                           0 & 0 & 0 & $-$ i
\end{array}
\right)
\left( \begin{array}{cccc} 0 & $-$1     & 0 & 0\\
                           1 & 0 & 0 & 0\\
                           0 & 0 & 0 & $-$1\\
                           0 & 0 & 1 & 0
\end{array}
\right)
\left( \begin{array}{cccc} 0 & 0     & $-$1 & 0\\
                           0 & 0 & 0 & $-$1\\
                           1 & 0 & 0 & 0\\
                           0 & 1 & 0 & 0
\end{array}
\right)$  & $=$ & 
$\left( \begin{array}{cccc} 0 & 0     & 0 & i\\
                           0 & 0 & i & 0\\
                           0 & $-$i & 0 & 0\\
                           $-$ i & 0 & 0 & 0
\end{array}
\right)~,$\\
$\gamma_{02}^l = - \gamma_{31}^l \gamma_5^l$ &
~~~$\leftrightarrow$ ~~~ &
$ - \left( \begin{array}{cccc} 0 & $-$1     & 0 & 0\\
                           1 & 0 & 0 & 0\\
                           0 & 0 & 0 & $-$1\\
                           0 & 0 & 1 & 0
\end{array}
\right)
\left( \begin{array}{cccc} 0 & 0     & $-$1 & 0\\
                           0 & 0 & 0 & $-$1\\
                           1 & 0 & 0 & 0\\
                           0 & 1 & 0 & 0
\end{array}
\right)$ & $=$ &
$\left( \begin{array}{cccc} 0 & 0     & 0 & $-$1\\
                           0 & 0 & 1 & 0\\
                           0 & 1 & 0 & 0\\
                           $-$ 1 & 0 & 0 & 0
\end{array}
\right)~,$\\
$\gamma_{03}^l = \gamma_{21}^l \gamma_5^l$ &
~~~$\leftrightarrow$ ~~~ &
$ \left( \begin{array}{cccc} i & 0     & 0 & 0\\
                           0 & $-$ i & 0 & 0\\
                           0 & 0 & i & 0\\
                           0 & 0 & 0 & $-$ i
\end{array}
\right)
\left( \begin{array}{cccc} 0 & $-$1     & 0 & 0\\
                           1 & 0 & 0 & 0\\
                           0 & 0 & 0 & $-$1\\
                           0 & 0 & 1 & 0
\end{array}
\right)$ & $=$ &
$\left( \begin{array}{cccc} 0 & 0     & $-$i & 0\\
                           0 & 0 & 0 & i\\
                           i & 0 & 0 & 0\\
                           0 & $-$i & 0 & 0
\end{array}
\right)~.$
\end{tabular}
\end{center}
The complete set of translation rules is given in Appendix A.

\section{The Dirac-Hestenes Lagrangian}

Our main objective in this work is to derive the Lagrangian,
${\cal L}_{H}$, which yields the Dirac-Hetsenes equation
\begin{equation}
\label{he2}
{\cal D}_+ \Psi_H \gamma_{21} = 
m \widehat{\Psi}_H~~~\in Cl_{1,3}^+ \otimes {\cal F} ~,
\end{equation}
\[
{\cal D}_{\pm} \equiv \partial_t \pm \gamma_{0 k}^l \partial_k ~,
~~~~~\mbox{\footnotesize $k=1,2,3$}~.
\]
We shall obtain the Dirac-Hestenes Lagrangian, ${\cal L}_{H}$, by translation.
To do that, let us start by considering the traditional form for the
complex Dirac Lagrangian,
\begin{equation}
\label{dl}
{\cal L}_D \equiv \Psi_D^{\dag} h \Phi_D~~~\in \C \otimes {\cal F}~,
~~~~~\Phi_D \equiv \left( i \gamma^{\mu} \partial_{\mu} - m \right) \Psi_D
~~~\in {\cal I} \otimes {\cal F}~.
\end{equation}
We showed~(\ref{l1}) that
\[
\C \otimes {\cal F} \ni ~~
\Psi^{\dag}_D h \Phi_D  ~~\leftrightarrow ~~ 
\chi \left( \widetilde{\Psi}_H \widehat{\Phi}_H \right)
~~ \in \D \otimes {\cal F}~,
\]
thus, to obtain the desired translation we need to calculate
\[
\widehat{\Phi}_H = \alpha ( \Phi_H ) =
\rho \left( \gamma_0 \Phi_D \right)~~\in Cl_{1,3}^+ \otimes {\cal F}~.
\]
By using the results presented in the previous Section, we find
\[
\rho \left( \gamma_0 \Phi_D \right) = 
\rho \left[ \left( i \gamma_0 \gamma^{\mu} \partial_{\mu} - m \gamma_0 
             \right) \Psi_D \right]
= {\cal D}_+ \Psi_H - m \widehat{\Psi}_H~,
\]
and consequently, 
\begin{equation}
\label{dhl}
\C \otimes {\cal F} \ni ~~
{\cal L}_D ~ \leftrightarrow ~ {\cal L}_{H} \equiv
\chi \left( \widetilde{\Psi}_H {\cal D}_+ \Psi_H \gamma_{21} - 
m \widetilde{\Psi}_H \widehat{\Psi}_H \right)
~~\in \D \otimes {\cal F}~.
\end{equation}

Let us now discuss the hermiticity of the Dirac-Hetsenes Lagrangian, 
${\cal L}_H$. By applying the reversion involution to ${\cal L}_H$
we get
\[
\widetilde{\cal L}_H = \chi 
\left( - \gamma_{21} \widetilde{\Psi}_H 
\stackrel{\leftarrow}{\cal D}_+ \Psi_H - 
m \overline{\Psi}_H \Psi_H \right)~,
\]
where $\stackrel{\leftarrow}{\cal D}_+$ indicates the left-action on 
$\widetilde{\Psi}_H$  of the derivation which appear in the operator 
${\cal D}_+$. By observing that
\[ \chi \left( \overline{\Psi}_H \Psi_H \right)  = 
\chi \left( \widetilde{\Psi}_H \widehat{\Psi}_H \right)~,
\]
and performing integration by parts, we obtain
\begin{equation}
\label{l2}
\widetilde{\cal L}_H = \chi 
\left( \gamma_{21} \widetilde{\Psi}_H 
{\cal D}_+ \Psi_H - 
m \widetilde{\Psi}_H \widehat{\Psi}_H \right)~.
\end{equation}
Due to the $\D $-complex geometry, 
the bivector $\gamma_{21}$ can be removed from the extreme left to
right $\widetilde{\Psi}_H {\cal D}_+ \Psi_H $ 
in Eq.~(\ref{l2}), and so
the hermiticity of the Dirac-Hestenes Lagrangian is proved,
\[ {\cal L}_H = \widetilde{\cal L}_H~. \]

In order to formulate the variational principle within the algebraic formalism,
let us rewrite Eq.~(\ref{dhl}), by using the projection operator 
\[ \mbox{End}_{\DD} \left( Cl_{1,3}^+ \right) \ni 
   {\cal P} \equiv \mbox{{\footnotesize $\frac{1}{2}$}} \, 
\left( 1 - \gamma_{21}^l\gamma_{21}^r \right)~, \]
and the grade-involution $\alpha$. The new expression for the Dirac-Hestenes
Lagrangian reads
\[ 
{\cal L}_H = \mbox{{\footnotesize $\frac{1}{2}$}} \, \left\{ {\cal P} 
\left( \widetilde{\Psi}_H {\cal D}_+ \Psi_H \gamma_{21} - 
m \widetilde{\Psi}_H \widehat{\Psi}_H \right) + 
\alpha \left[
{\cal P} 
\left( \widetilde{\Psi}_H {\cal D}_+ \Psi_H \gamma_{21} - 
m \widetilde{\Psi}_H \widehat{\Psi}_H \right) 
\right] \right\}
~~~\in \D \otimes {\cal F} ~,
\]
or by expliciting the action of the $\cal P$-operator
and $\alpha$-involution,
\begin{eqnarray}
\label{lag}
{\cal L}_H & ~=~ \frac{1}{4} \, ( \, &  
                          \widetilde{\Psi}_H {\cal D}_+ \Psi_H \gamma_{21} 
                          - m \widetilde{\Psi}_H \widehat{\Psi}_H +
\nonumber \\
       &  & \gamma_{21} \widetilde{\Psi}_H {\cal D}_+ \Psi_H 
         + m \gamma_{21}  \widetilde{\Psi}_H \widehat{\Psi}_H \gamma_{21}+
\nonumber \\
       &  & \overline{\Psi}_H {\cal D}_- \widehat{\Psi}_H \gamma_{21} 
         - m \overline{\Psi}_H \Psi_H +
\nonumber \\
       &  & \gamma_{21} \overline{\Psi}_H {\cal D}_- \widehat{\Psi}_H 
         + m \gamma_{21}  \overline{\Psi}_H \Psi_H \gamma_{21} \, )~.
\end{eqnarray}
It is here that appeal to the variational principle must be made. A variation
$\delta \Psi_H$ in $\Psi_H$ from Eq.~(\ref{lag}) cannot be brought to the 
extreme right because of the bivector $\gamma_{21}$ in the first term of 
the previous expression. The only consistent procedure is to generalize the 
variational rule that says that $\Psi_H$ and $\overline{\Psi}_H$ must be 
varied {\em independently}~\cite{DRL}. 
We thus apply independent variations to
\begin{equation}
\label{e11}
\Psi_H ~,~~ \Psi_H \gamma_{21} ~,~~ 
\widehat{\Psi}_H ~,~~ \widehat{\Psi}_H \gamma_{21}~,
\end{equation} 
and
\begin{equation}
\label{e22}
\overline{\Psi}_H ~,~~ \gamma_{21} \overline{\Psi}_H  ~,~~ 
\widetilde{\Psi}_H ~,~~ \gamma_{21} \widetilde{\Psi}_H ~.
\end{equation}
This generalization of the variational principle is discussed in appendix~B.
The variations applied to fields~(\ref{e11}) yield the adjoint
Dirac-Hestenes equation
\begin{equation}
\label{ea}
- \gamma_{21} \widetilde{\Psi}_H 
\stackrel{\leftarrow}{\cal D}_+ = m \overline{\Psi}_H~,
\end{equation}
whereas that ones applied to fields~(\ref{e22}) yield the 
Dirac-Hestenes equation
\begin{equation}
\label{e2}
{\cal D}_+ \Psi_H \gamma_{21} = 
m \widetilde{\Psi}_H ~.
\end{equation}

Let us discuss a interesting point. The Dirac-Hestenes Lagrangian~(\ref{dhl})
is $\D $-complex and hermitian. 
The situation is more subtle with classical field for
now ${\cal L}_H^{new}$, defined by 
\[  
{\cal L}_H^{new} = \mbox{{\footnotesize $\frac{1}{2}$}} \, \left( \,   
\widetilde{\Psi}_H {\cal D}_+ \Psi_H \gamma_{21} + 
\gamma_{21} \widetilde{\Psi}_H {\cal D}_+ \Psi_H 
- m \widetilde{\Psi}_H \widehat{\Psi}_H 
- m \overline{\Psi}_H \Psi_H \, \right)~~\in {\cal F}~, 
\]
is both hermitian and {\em real}. Thus it may be objected that the complex 
projection in the previous classical Lagrangian is superfluous. For 
${\cal L}_H^{new}$ itself this true but for multivectorial algebraic 
variations in the fields, $\delta \Psi_H$, etc., a difference exists.
The variation $\delta {\cal L}_H^{new} \in Cl_{1,3}^+ \otimes {\cal F}$,  
while $\delta {\cal L}_H$ from~(\ref{dhl}) is always $\D $-complex.  
Furthermore, ${\cal L}_H^{new}$ does not yield the correct field equation 
through the variational principle unless we limit $\delta \Psi_H$, etc., to
$\D $-complex variations notwithstanding 
$\Psi_H \in Cl_{1,3}^+ \otimes {\cal F}$. We consider this latter option 
unjustified and thus select for the formal structure of the classical
Lagrangian that of Eq.~(\ref{dhl}). Let us summarize the situation 
concerning fields and variations
\begin{eqnarray*}
\Psi_H \, , ~ \widehat{\Psi}_H 
\, , ~ \widehat{\Psi}_H 
\, , ~ \overline{\Psi}_H
\, , ~ \widetilde{\Psi}_H 
& ~~ \in ~~ & Cl_{1,3}^+ \otimes {\cal F}~,\\
\delta (\Psi_H) \, , ~ \delta (\Psi_H \gamma_{21}) \, ,~ 
\delta (\widehat{\Psi}_H) \, ,~ \delta ( \widehat{\Psi}_H \gamma_{21}) 
& ~~ \in ~~ & Cl_{1,3}^+ \otimes {\cal F}~,\\
\delta ( \overline{\Psi}_H ) \, , ~ \delta( \gamma_{21} \overline{\Psi}_H ) 
\, , ~ \delta (\widetilde{\Psi}_H) 
\, , ~ \delta (\gamma_{21} \widetilde{\Psi}_H) 
& ~~ \in ~~ & Cl_{1,3}^+ \otimes {\cal F}~,\\
{\cal L}_H 
& ~~ \in ~~ & \D \otimes {\cal F}~,\\
\delta {\cal L}_H 
& ~~ \in ~~ & \D  \otimes {\cal F}~.
\end{eqnarray*}

We conclude this Section by discussing an alternative way to obtain the 
field equations from the Dirac-Hestenes Lagrangian. In doing that, let us 
rewrite the $\alpha$-involution by using the operator 
$\gamma_0^l \gamma_0^r$
\[ \alpha \left( \Psi_H \right) \equiv \widehat{\Psi}_H = 
\gamma_0 \Psi_H \gamma_0~~\in Cl_{1,3}^+ \otimes {\cal F}~.
\]
By adopting this notation we can express the Dirac-Hestenes Lagrangian as
\begin{equation}
\label{l11}
{\cal L}_H = {\cal P} {\cal P}_{\alpha} 
\left[
\widetilde{\Psi}_H {\cal D}_+ \Psi_H \gamma_{21} - 
m \widetilde{\Psi}_H \gamma_0 \Psi_H \gamma_0 
\right]~~\in \D \otimes {\cal F}~,
\end{equation}
where
\[ {\cal P}_{\alpha} \equiv \mbox{{\footnotesize $\frac{1}{2}$}} \, 
\left( 1 + \gamma_0^l \gamma_0^r \right)~,\]
\[ \left[ {\cal P} \, , \, {\cal P}_{\alpha} \right] = 0~. \]
In making the variation
\begin{equation}
\label{var}
\Psi_H \rightarrow \Psi_H + \delta \Psi_H~,
\end{equation}
we can put $\delta \Psi_H$ on the extreme right because, due to our mapping on
$\D \otimes {\cal F}$, we can bring $\gamma_{21}$ and $\gamma_0$ from
the extreme right to left in Eq.~(\ref{l11}). In fact,
\begin{eqnarray*}
{\cal P} \left( {\cal A} \gamma_{21} \right) & = & 
{\cal P} \left( \gamma_{21} {\cal A} \right)~,\\
{\cal P}_{\alpha} \left( {\cal A} \gamma_{0} \right) & = & 
{\cal P}_{\alpha} \left( \gamma_0 {\cal A}  \right)~,
\end{eqnarray*}
with
\[ {\cal A} \in Cl_{1,3} \otimes {\cal F}~.\]
The variation~(\ref{var}) implies
\[
\delta {\cal L}_H = {\cal P} {\cal P}_{\alpha} 
\left[ 
\widetilde{\Psi}_H {\cal D}_+ \delta \Psi_H \gamma_{21} - 
m \widetilde{\Psi}_H \gamma_0 \delta \Psi_H \gamma_0 
\right]~,   
\]
which after integration by parts and by moving $\gamma_{21}$ and $\gamma_0$
from the extreme right to left, becomes
\[
\delta {\cal L}_H = {\cal P} {\cal P}_{\alpha} 
\left[
- \gamma_{21} \widetilde{\Psi}_H 
\stackrel{\leftarrow}{\cal D}_+ \delta \Psi_H - 
m \gamma_0 \widetilde{\Psi}_H \gamma_0 \delta \Psi_H 
\right]~.
\]
Finally, $\delta {\cal L}_H = 0$ implies
\[
- \gamma_{21} \widetilde{\Psi}_H 
\stackrel{\leftarrow}{\cal D}_+ = 
m \gamma_0 \widetilde{\Psi}_H \gamma_0 ~, 
\]
and so we obtain the adjoint Dirac-Hestenes equation~(\ref{ea}), 
as required.

\section{The Invariance Group of ${\cal L}_{H}$}

Having obtained the Dirac-Hestenes Lagrangian in the previous Section, we 
may ask which global group leaves this Lagrangian invariant. Remembering that
$\Psi_H \in Cl_{1,3}^+ \otimes {\cal F}$, the most general $\D$-complex linear
transformation on $\Psi_H$ is given by
\begin{equation}
\label{it}
\Psi_H ~ \rightarrow ~ \left( A^l + B^l \gamma_{21}^r + 
C^l \gamma_0^l \gamma_0^r + D^l \gamma_{21}^r \gamma_{0}^l \gamma_{0}^r
\right) \Psi_H ~~\in Cl_{1,3}^+ \otimes {\cal F}~,
\end{equation}
with
\[ 
A^l \, , \, B^l \, , \,  C^l \, , \, D^l \, \in Cl_{1,3}^{+ \, (l)}~. 
\]
Now, the algebraic structure of the Dirac operator ${\cal D}_+$ strongly 
limits the left action on $\Psi_H$, this leads to the conclusion that
\[ 
A^l = a \cdot 1^l \, , \, B^l = b \cdot 1^l \, , \,  
C^l = 0  \, , \, D^l = 0 ~,
~~~~~a,b \in \R ~. 
\]
So, Eq.~(\ref{it}) will be modified as
\begin{equation}
\label{it2}
\Psi_H ~ \rightarrow ~ \left( a \cdot 1^l + b \cdot \gamma_{21}^r \right) 
\Psi_H
\equiv \Psi_H \,  ( a + \gamma_{21} b )~,
\end{equation}
and consequently
\begin{eqnarray}
\label{it3}
\widehat{\Psi}_H & ~\rightarrow ~ & 
\widehat{\Psi}_H \, ( a + \gamma_{21} b )~, \nonumber \\
\widetilde{\Psi}_H & ~\rightarrow ~ & 
( a - \gamma_{21} b ) \,  \widetilde{\Psi}_H ~.
\end{eqnarray}
Applying the global transformations~(\ref{it2}-\ref{it3}), the Dirac-Hestenes 
Lagrangian becomes
\begin{eqnarray*}
{\cal L}'_{H} & \equiv & 
\chi \left[ z^* \left( \widetilde{\Psi}_H {\cal D}_+ \Psi_H \gamma_{21} - 
m \widetilde{\Psi}_H \widehat{\Psi}_H \right) z \right]\\
              & \equiv & z^* z  ~ 
\chi \left( \widetilde{\Psi}_H {\cal D}_+ \Psi_H \gamma_{21} - 
m \widetilde{\Psi}_H \widehat{\Psi}_H \right)~,\\
 & & z , z^* \in \D ~.
\end{eqnarray*}
Thus by requiring $z^* z =1$, we find that the only invariance group is 
defined by 
\[ U(1,\gamma_{21}^r) ~, \]
where the previous notation means the right action of the $\D$-complex 
unitary group on the algebraic spinor $\Psi_H$
\begin{equation}
\label{ra}
\Psi_H ~ \rightarrow ~ e^{\gamma_{21}^r \delta} \Psi_H \equiv
                       \Psi_H \, e^{\gamma_{21} \delta}~,   
~~~~~\delta \in \R ~.
\end{equation}

Remembering that the Glashow group~\cite{GLA} for the 
Salam-Weinberg theory~\cite{S1,W2} 
is $SU(2) \otimes U(1)$, we observe that this 
$U(1)$ group may be identified with ours $U(1,\gamma_{21}^r)$ and our field
$\Psi_H \in Cl_{1,3}^+ \otimes {\cal F}$ must necessarily be a singlet (scalar)
under $SU(2)$.

The interesting feature is what happens if we select a field in the full
Space Time algebra $Cl_{1,3} \otimes {\cal F}$. Now the number of fermionic 
particles is two
\[ \Psi_H^{(1)} +  \Psi_H^{(2)} \gamma_0 ~,~~~~~ 
\Psi_H^{(1,2)} \in   Cl_{1,3}^+ \otimes {\cal F}~.
\]
For example the leptons of the first family (electronic neutrino $\nu_e$,
electron $e$) can be concisely rewritten in $Cl_{1,3} \otimes {\cal F}$ as
\begin{equation}
\Psi_{Lep}^{(1^{st} \, fam)} = 
\Psi_H^{(\nu_e)} +  \Psi_H^{(e)} \gamma_0 ~,~~~~~ 
\Psi_H^{(\nu_e , e)} \in   Cl_{1,3}^+ \otimes {\cal F}~.
\end{equation}
The orthogonality of the fields $\Psi_H^{(\nu_e)}$, $\Psi_H^{(e)} \gamma_0 $
is guaranteed by our $\D $-complex mapping,
\[
\chi \left( \widetilde{\Phi} \widehat{\Psi} \gamma_0 \right) = 0~,
~~~~~\Phi \, , \, \Psi \, \in   Cl_{1,3}^+ \otimes {\cal F}~.
\]
Now it is still not obvious, due to the presence of the Dirac operator
${\cal D}_+$, that an invariance group isomorphic to $SU(2)$ exists.
We remark that to obtain a global invariance
isomorphic to $SU(2)$ we must choose suitable combinations of
\[ \gamma_5^{l,r} \, , \, \gamma_0^r \, , \, \gamma_{21}^r~.\]
These operators satisfy 
\[ \left[ \gamma_5^l \, , \, {\cal D}_+ \right] = 0~, \]
and
\[ 
\chi \left( \left[ {\cal A} \, , \, \gamma_5 \right] \right) =
\chi \left( \left[ {\cal A} \, , \, \gamma_0 \right] \right) =
\chi \left( \left[ {\cal A} \, , \, \gamma_{21} \right] \right) = 0~,
~~~~~{\cal A} \in   Cl_{1,3} \otimes {\cal F}~.
\]
Consequently, the following infinitesimal transformation
\[ \Psi_{Lep} ~\rightarrow ~ \left( 
1 + \alpha_1 \gamma_0^r \gamma_{21}^r  + 
    \alpha_2 \gamma_5^l \gamma_5^r \gamma_0^r + 
    \alpha_3 \gamma_5^l \gamma_5^r \gamma_{21}^r + 
    \beta \gamma_{21}^r \right) \Psi_{Lep}~,~~~~~~
1 \gg \alpha_{1,2,3} \, , \, \beta \in \R~,
\]
leaves invariant the zero-mass Lagrangian
\begin{equation}
{\cal L}_{Lep} \equiv 
\chi \left( \widetilde{\Psi}_{Lep} {\cal D}_+ \Psi_{Lep} \gamma_{21} \right)
~~\in \D \otimes {\cal F}~.
\end{equation}
The zero-mass fields will gain mass by spontaneous symmetry 
breaking~\cite{SB1,SB2}.

The antihermitian generators
\[ \gamma_0^r \gamma_{21}^r ~ , ~ 
   \gamma_5^l \gamma_5^r \gamma_0^r ~ , ~
   \gamma_5^l \gamma_5^r \gamma_{21}^r ~~~\mbox{and}~~~ 
   \gamma_{21}^r ~
\]
represent the multivectorial $Cl_{1,3}$-counterpart of the generators of the
standard (complex) Glashow group 
\[ 
SU(2) \otimes U(1)~.
\]

\section{Conclusions}

We begin our discussion from the end results of the last Section. We have
shown that by working within a multivectorial formalism it is possible to
impose a Glashow group invariance and that this occurs by merely adopting
$Cl_{1,3}$-fields. Our viewpoint is that the $SU(2) \otimes U(1)$ invariance 
in Particle Physics could be better understood by working in the 
$Cl_{1,3}$-formalism, where each element is suitable of 
{\em geometric interpretations}. For example, a better understanding of the 
geometric meaning of the generators of the invariance Glashow group could be
very important in reaching grand unification groups. The adoption of a
$\D $-complex geometry represents a fundamental ingredient of the 
multivectorial algebraic approach to Quantum Mechanics. Such a mapping gives 
the desired {\em electromagnetic invariance} $U(1,\gamma_{21}^r)$ and suggests
an invariance group isomorphic to the Glashow group. By passing from 
$Cl_{1,3}^+$ to $Cl_{1,3}$ fields, the $\D $-complex geometry guarantees the
right orthogonality between electron and neutrino field and gives the 
possibility to find four $Cl_{1,3}^{(l/r)}$-elements which are isomorphic
to the generators of the electroweak group $SU(2) \otimes U(1)$.
A complete discussion on the Salam-Weinberg model in the multivectorial
formalism will be presented in a forthcoming paper~\cite{swm}.

Let us recall the other result of this paper. We discussed and generalized 
the application of the variational principle to Lagrangians with 
$Cl_{1,3}^+$-fields. In order to obtain the Dirac-Hestenes equation we proved 
the need to adopt a $\D $-complex mapping for our Lagrangians or apply,
due to the noncommutative nature of the Clifford algebras, different 
variations for the fields $\Psi_H$, $\Psi_H \gamma_{21}$, etc.

We also recall the possibility to perform a {\em translation} between 
$4 \times 4$ complex matrices and left/right elements of the even 
Space Time sub-algebra. This allows an immediate translation of the Dirac
equation in the multivectorial formalism. Obviously this approach can be used
to reproduce others standard results of Quantum Mechanics. 
We conclude emphasizing that this translation represents only a 
{\em partial translation}, for example it does not apply to odd-dimensional 
complex matrices.  Different outputs can be obtained by working with 
Clifford Algebras. {\em New geometric interpretations} naturally appear in 
the Space Time Algebraic approach and this could be very useful in 
reaching fundamental symmetries in unification Lagrangians.


\acknowledgements

The authors {\it S.d.L.} and {\it Z.O.}  wish to express their thanks to 
IMECC, University of Campinas, where the paper was written, for financial 
support. {\it S.d.L.} acknowledges the many helpful suggestions and 
comments of S.~Adler and P.~Rotelli during the preparation of the paper and 
he is greatly indebted to the Brazilian colleagues and friends for their warm 
hospitality.


\newpage

\section*{Appendix A. Translation Rules}

Let define projectors
\[ 
\alpha_{\pm} \equiv \mbox{\footnotesize $\frac{1}{2}$} 
\, ( \mbox{id} \pm \alpha ) 
\in \mbox{End}_{\DD} \left( Cl_{1,3}^+ \right)~.
\]
The 16 linear independent $4\times 4$ matrices have the following counterparts
in the even Space Time sub-Algebra 
{\footnotesize 
\begin{center}
\begin{tabular}{rcccrcccrcccrccc}
$\alpha_+$ & $\leftrightarrow$ & 
$\left( \begin{array}{cccc} 1 &   0 & 0 & 0\\
                            0 & 1 & 0 & 0\\
                            0 & 0 & 0 & 0\\
                            0 & 0 & 0 & 0
\end{array}
\right)$ & , & 
~$\gamma_{21}^l \gamma_{21}^r \alpha_+$ & $\leftrightarrow$ & 
$\left( \begin{array}{cccc} $-$ 1 &   0 & 0 & 0\\
                           0 & 1 & 0 & 0\\
                           0 & 0 & 0 & 0\\
                           0 & 0 & 0 & 0
\end{array}
\right)$ & , & 
~$\gamma_{23}^l \gamma_{21}^r \alpha_+$ & $\leftrightarrow$ & 
$\left( \begin{array}{cccc} 0 &   1 & 0 & 0\\
                           1 & 0 & 0 & 0\\
                           0 & 0 & 0 & 0\\
                           0 & 0 & 0 & 0
\end{array}
\right)$ & , & 
~$\gamma_{31}^l \alpha_+$ & $\leftrightarrow$ & 
$\left( \begin{array}{cccc} 0 &   $-$ 1 & 0 & 0\\
                           1 & 0 & 0 & 0\\
                           0 & 0 & 0 & 0\\
                           0 & 0 & 0 & 0
\end{array}
\right)$ & , \\
$\alpha_-$ & $\leftrightarrow$ & 
$\left( \begin{array}{cccc} 0 &   0 & 0 & 0\\
                            0 & 0 & 0 & 0\\
                            0 & 0 & 1 & 0\\
                            0 & 0 & 0 & 1
\end{array}
\right)$ & , & 
~$\gamma_{21}^l \gamma_{21}^r \alpha_-$ & $\leftrightarrow$ & 
$\left( \begin{array}{cccc} 0 &   0 & 0 & 0\\
                           0 & 0 & 0 & 0\\
                           0 & 0 & $-$ 1 & 0\\
                           0 & 0 & 0 & 1
\end{array}
\right)$ & , & 
~$\gamma_{23}^l \gamma_{21}^r \alpha_-$ & $\leftrightarrow$ & 
$\left( \begin{array}{cccc} 0 &   0 & 0 & 0\\
                           0 & 0 & 0 & 0\\
                           0 & 0 & 0 & 1\\
                           0 & 0 & 1 & 0
\end{array}
\right)$ & , & 
~$\gamma_{31}^l \alpha_-$ & $\leftrightarrow$ & 
$\left( \begin{array}{cccc} 0 &   0 & 0 & 0\\
                           0 & 0 & 0 & 0\\
                           0 & 0 & 0 & $-$ 1\\
                           0 & 0 & 1 & 0
\end{array}
\right)$ & , \\
$- \gamma_5^l \alpha_+$ & $\leftrightarrow$ & 
$\left( \begin{array}{cccc} 0 &   0 & 1 & 0\\
                           0 & 0 & 0 & 1\\
                           0 & 0 & 0 & 0\\
                           0 & 0 & 0 & 0
\end{array}
\right)$ & , & 
~$\gamma_{03}^l \gamma_{21}^r \alpha_+$ & $\leftrightarrow$ & 
$\left( \begin{array}{cccc} 0 &   0 & 1 & 0\\
                           0 & 0 & 0 & $-$ 1\\
                           0 & 0 & 0 & 0\\
                           0 & 0 & 0 & 0
\end{array}
\right)$ & , & 
~$-\gamma_{01}^l \gamma_{21}^r \alpha_+$ & $\leftrightarrow$ & 
$\left( \begin{array}{cccc} 0 &   0 & 0 & 1\\
                           0 & 0 & 1 & 0\\
                           0 & 0 & 0 & 0\\
                           0 & 0 & 0 & 0
\end{array}
\right)$ & , &
~$\gamma_{02}^l \alpha_+$ & $\leftrightarrow$ & 
$\left( \begin{array}{cccc} 0 &   0 & 0 & $-$ 1\\
                           0 & 0 & 1 & 0\\
                           0 & 0 & 0 & 0\\
                           0 & 0 & 0 & 0
\end{array}
\right)$ & , \\
$\gamma_5^l \alpha_-$ & $\leftrightarrow$ & 
$\left( \begin{array}{cccc} 0 &   0 & 0 & 0\\
                           0 & 0 & 0 & 0\\
                           1 & 0 & 0 & 0\\
                           0 & 1 & 0 & 0
\end{array}
\right)$ & , & 
~$\gamma_{03}^l \gamma_{21}^r \alpha_-$ & $\leftrightarrow$ & 
$\left( \begin{array}{cccc} 0 &   0 & 0 & 0\\
                           0 & 0 & 0 & 0\\
                           $-$ 1 & 0 & 0 & 0\\
                           0 & 1 & 0 & 0
\end{array}
\right)$ & , & 
~$\gamma_{01}^l \gamma_{21}^r \alpha_-$ & $\leftrightarrow$ & 
$\left( \begin{array}{cccc} 0 &   0 & 0 & 0\\
                           0 & 0 & 0 & 0\\
                           0 & 1 & 0 & 0\\
                           1 & 0 & 0 & 0
\end{array}
\right)$ & , &
~$\gamma_{02}^l \alpha_-$ & $\leftrightarrow$ & 
$\left( \begin{array}{cccc} 0 &   0 & 0 & 0\\
                           0 & 0 & 0 & 0\\
                           0 & 1 & 0 & 0\\
                           $-$ 1 & 0 & 0 & 0
\end{array}
\right)$ & .  
\end{tabular}
\end{center}
}
The remaining 16 ``complex'' matrices are obtained by 
$\gamma_{21}^r \leftrightarrow i \openone_4$ multiplication. The 16 operators 
\[ \left( 1^l \, , ~\gamma_{21}^l \gamma_{21}^r \, ,  
                 ~\gamma_{23}^l \gamma_{21}^r \, ,  
                 ~\gamma_{31}^l \, ,  
                 ~\gamma_{5}^l  \, ,  
                 ~\gamma_{03}^l \gamma_{21}^r \, ,  
                 ~\gamma_{02}^l \, ,  
                 ~\gamma_{01}^l \gamma_{21}^r \right) \alpha_{\pm}~~~
\in \mbox{End}_{\DD} \left( Cl_{1,3}^+ \right)~,
\]
are $\D $-complex linear independent,
\[ 
\mbox{dim}_{\DD} Cl_{1,3}^{+ \, (l/r)} = 
\mbox{dim}_{\CC } \mbox{Mat}_4( \C ) = 16~.
\]
The proof is based on the $i$-complex linear independence of the listed
$4 \times 4$ real matrices.

\section*{Appendix B. Variational Principle}

Consider one of the simplest of all particle Lagrangian densities, that for 
two classical scalar fields, $\varphi_{1,2} \in {\cal F}$, without interactions
\begin{eqnarray}
{\cal L} &    =   & 
\mbox{\footnotesize{$\frac{1}{2}$}} \,  \partial_{\mu} \varphi_{1} 
\partial^{\mu} \varphi_{1} - 
\mbox{\footnotesize{$\frac{m^{2}}{2}$}} \, \varphi_{1}^{\; 2} + 
\mbox{\footnotesize{$\frac{1}{2}$}} \,  \partial_{\mu} \varphi_{2} 
\partial^{\mu} \varphi_{2} - 
\mbox{\footnotesize{$\frac{m^{2}}{2}$}} \, \varphi_{2}^{\; 2} \nonumber \\
         & \equiv & \partial_{\mu} \phi^{\dag} \partial^{\mu} \phi - 
m^{2} \phi^{\dag} \phi  
\end{eqnarray}
where 
\[ 
\phi \equiv \mbox{\footnotesize $\frac{1}{\sqrt{2}}$} \,
( \varphi_{1} + i \varphi_{2} ) ~~
\in \C \otimes {\cal F}~,~~  
\phi^{\dag} \equiv \mbox{\footnotesize $\frac{1}{\sqrt{2}}$} \,
( \varphi_{1} - i \varphi_{2} ) ~~~
\in \C \otimes {\cal F}~.
\]
The well known corresponding Euler-Lagrange equations are:
\begin{equation}
\left( \partial_{\mu} \partial^{\mu} + m^{2} \right) \varphi_{1,2} = 0 ~,
\end{equation}
or, equivalently,
\begin{equation}
\left( \partial_{\mu} \partial^{\mu} + m^{2} \right) \phi = 0 ~.
\end{equation}
Now to obtain ``directly'' the last equation one performs very particular 
variations of $\phi$ and $\phi^{\dag}$
\begin{eqnarray}
\label{five}
\phi  \, ~     & \rightarrow & \phi \nonumber \\
\phi^{\dag}   & \rightarrow & \phi^{\dag} + \delta \phi^{\dag}
\end{eqnarray}
i.e. in order to obtain the corresponding Euler-Lagrangian equation one 
treats $\phi$ and $\phi^{\dag}$ as {\em independent} fields. In second 
quantization these fields indeed contain independent creation and 
annihilation operators corresponding to positive and negative charged 
particles. To satisfy Eq.~(\ref{five}) we must necessarily have,
\begin{equation}
\delta \varphi_{1} + i \delta \varphi_{2} = 0
\end{equation}
and this means, that the {\em variations} in the originally real 
$\varphi_{1,2}$ fields are complex (if $\delta \varphi_{1}$ is real then 
$\delta \varphi_{2}$ is pure imaginary etc.).

In this Appendix we aim to generalize the variational rule given for 
``complex'' fields. Let $\Psi \in Cl_{1,3}^+ \otimes {\cal F}$ be expressed
by
\[ 
\Psi = \psi_0 + 
\gamma_{01} \psi_1 +  \gamma_{02} \psi_2 +  \gamma_{03} \psi_3 + 
\gamma_{21} \psi_4 +  \gamma_{31} \psi_5 +  \gamma_{23} \psi_6 + 
\gamma_{5} \psi_7~,~~~~~\psi_{0,...,7} \in {\cal F}~.
\]
As shown in the Introduction, we can define the involutions
\begin{eqnarray*}
\widehat{\Psi} & = & 
\psi_0 -
\gamma_{01} \psi_1 -  \gamma_{02} \psi_2 -  \gamma_{03} \psi_3 + 
\gamma_{21} \psi_4 +  \gamma_{31} \psi_5 +  \gamma_{23} \psi_6 -
\gamma_{5} \psi_7~,\\
\widetilde{\Psi} & = & 
\psi_0 + 
\gamma_{01} \psi_1 +  \gamma_{02} \psi_2 +  \gamma_{03} \psi_3 - 
\gamma_{21} \psi_4 -  \gamma_{31} \psi_5 -  \gamma_{23} \psi_6 - 
\gamma_{5} \psi_7~,\\
\overline{\Psi} & = & 
\psi_0 -
\gamma_{01} \psi_1 -  \gamma_{02} \psi_2 -  \gamma_{03} \psi_3 - 
\gamma_{21} \psi_4 -  \gamma_{31} \psi_5 -  \gamma_{23} \psi_6 + 
\gamma_{5} \psi_7~.
\end{eqnarray*}
The complex variational principle which treats $\Phi$ and $\Phi^{\dag}$
as independent fields is now generalized by 
applying different variations to $\Psi$, $\widehat{\Psi}$,
$\widetilde{\Psi}$, $\overline{\Psi}$. Nevertheless, by working 
within the noncommutative algebra $Cl_{1,3}^+$ we must also analyze the 
following fields
\begin{center}
\begin{tabular}{ccc}
$-\gamma_{21} \Psi \gamma_{21}~,$ &
~~$-\gamma_{31} \Psi \gamma_{31}~,$~~&
~~$-\gamma_{23} \Psi \gamma_{23}~,$~~\\
$-\gamma_{21} \widehat{\Psi} \gamma_{21}~,$ &
~~$-\gamma_{31} \widehat{\Psi} \gamma_{31}~,$~~&
~~$-\gamma_{23} \widehat{\Psi} \gamma_{23}~,$~~\\
$-\gamma_{21} \widetilde{\Psi} \gamma_{21}~,$ &
~~$-\gamma_{31} \widetilde{\Psi} \gamma_{31}~,$~~&
~~$-\gamma_{23} \widetilde{\Psi} \gamma_{23}~.$~~
\end{tabular}
\end{center}
In fact, we can treat $\Psi$ and $\Phi = -\gamma_{21} \Psi \gamma_{21}$ as 
independent fields
\begin{eqnarray*}
\Psi ~& \rightarrow &~ \Psi~,\\
\Phi ~& \rightarrow &~ \Phi + \delta \Phi~.
\end{eqnarray*}
The previous equation is satisfied by requiring
\[
\delta \psi_0 + 
\gamma_{01} \delta \psi_1 +  \gamma_{02} \delta \psi_2 +  
\gamma_{03} \delta \psi_3 + 
\gamma_{21} \delta \psi_4 +  \gamma_{31} \delta \psi_5 +  
\gamma_{23} \delta \psi_6 + 
\gamma_{5} \delta \psi_7 = 0~,
\]
and this means that the variations in the originally real fields 
$\psi_{0,...,7}$ are in $Cl_{1,3}^+$. In conclusion, we must apply
different variations to the fields
\[
\Psi~,~ \widehat{\Psi}~,~ \widetilde{\Psi}~,~ \overline{\Psi}~,~
\Phi~,~ \widehat{\Phi}~,~ \widetilde{\Phi}~,~ \overline{\Phi}~,
\]
which appear in the Dirac-Hestenes Lagrangian~(\ref{lag})
\begin{eqnarray}
{\cal L}_H & ~=~ \frac{1}{4} \, ( \, &  
                          \widetilde{\Psi}_H {\cal D}_+ \gamma_{21} \Phi_H 
                          - m \widetilde{\Psi}_H \widehat{\Psi}_H +
\nonumber \\
       &  & \widetilde{\Phi}_H \gamma_{21} {\cal D}_+ \Psi_H 
         - m \widetilde{\Phi}_H \widehat{\Phi}_H +
\nonumber \\
       &  & \overline{\Psi}_H {\cal D}_- \gamma_{21} \widehat{\Phi}_H  
         - m \overline{\Psi}_H \Psi_H +
\nonumber \\
       &  & \overline{\Phi}_H \gamma_{21} {\cal D}_- \widehat{\Psi}_H 
         - m \overline{\Phi}_H \Phi_H \, )~.
\end{eqnarray}



\begin{references}

\bibitem{DIR}
P.~A.~M.~Dirac, 
   {\em The Quantum Theory of the Electron},
   Proc.~Roy.~Soc.~of London {\bf A117}, 610 (1928).
\bibitem{ZB}
J.~D.~Bjorken and S.~D.~Drell,
   {\em Relativistic Quantum Mechanics} (McGraw-Hill, New York, 1964).\\
C.~Itzykson and J.~B.~Zuber,
   {\em Quantum Field Theory} (McGraw-Hill, New York, 1985).
\bibitem{HESP}
D.~Hestenes, 
   J.~Math.~Phys.~{\bf 8}, 798 (1967), {\em ibidem} {\bf 16}, 556 (1975); 
   Phys.~Teach.~{\bf 17}, 235 (1979);  Found.~Phys.~{\bf 20}, 1213 (1990).
\bibitem{SS}
A.~Sommerfeld, {\em Atombau und Spektrallinien Vieweg} 
   (Braunschweig, Berlin, 1942).           
\bibitem{CA1}
D.~Hestenes, 
   {\em Space-Time Algebra}
   (Gordon \& Breach, New York, 1966).\\
D.~Hestenes and G.~Sobczyk, 
   {\em Clifford Algebra to Geometric Calculus}
   (D.~Riedel Publishing Company, Dordrecht, 1984).\\
D.~Hestenes and A.~Weingartshofer, 
   {\em The Electron, New Theory and Experiment}
   (Kluwer Academic Publishers, Dordrecht, 1991).
\bibitem{CA2}
P.~Lounesto,
   {\em Clifford Algebras and Spinors}
   (Cambridge UP, Cambridge, 1997).
\bibitem{CA3}
P.~Lounesto, 
   in P.~Letelier and W.~A.~Rodrigues (eds.),  
   {\em Gravitation}: {\em The Space-Time Structure}
   (World Scientific, Singapore, 1994), p.~50;
   Found.~Phys.~{\bf 16}, 967 (1986); 
                {\bf 23}, 1203 (1993).
\bibitem{CA4}
J.~Keller,
   Adv.~in Appl.~Cliff.~Alg.~{\bf 3}, 147 (1993).
\bibitem{CA5}
S.~Gull, A.~Lasenby and C.~Doran,
   Found.~Phys.~{\bf 23}, 1175 (1993);
   {\em ibidem}, 1239 (1993).
\bibitem{ZEN}
J.~R.~Zeni, 
   in P.~Letelier and W.~A.~Rodrigues (eds.),  
   {\em Gravitation}: {\em The Space-Time Structure}
   (World Scientific, Singapore, 1994), p.~544.
\bibitem{REM}
J.~Rembieli\'nski,
   J.~Phys.~A {\bf 11}, 2323 (1978).
\bibitem{HB}
L.~P.~Horwitz and L.~C.~Biedenharn,
   Ann.~Phys.~{\bf 157}, 432 (1984).
\bibitem{DR}
S.~De Leo and W.~A.~Rodrigues, Int.~J.~Theor.~Phys.~{\bf 36}, 2725 (1997).\\
S.~De Leo and W.~A.~Rodrigues,  
   {\em Quaternionic Electron Theory: I-Dirac's Equation} and   
   {\em II-Geometry, Algebra and Dirac's Spinors},  
   Int.~J.~Theor.~Phys. (to be published in May 98).  
\bibitem{DRV}
S.~De Leo, W.~A.~Rodrigues and J.~Vaz,  
   {\em Complex Geometry and Dirac Equation} 
   (submitted for publication in {\em IJTP}).    
\bibitem{MAG}
M.~Gusiew-Czudzak and J.~Keller,
   Adv.~Appl.~Cliff.~Alg.~{\bf 7}, 419 (1997).
\bibitem{CRU}
A.~Crumeyrolle, 
   {\em Orthogonal and Symplectic Clifford Algebras. Spinor Structures}
   (Kluwer Academic Publishers, Dordrecht, 1990), 
   Mathematics and its Applications, vol.~57.
\bibitem{ADL}
S.~L.~Adler, 
   {\em Quaternion Quantum Mechanics and Quantum Field}
   (Oxford UP, New York, 1995).
\bibitem{DRL}
S.~De Leo and P.~Rotelli, 
   Mod.~Phys.~Lett.~A {\bf 11}, 357 (1996).
\bibitem{GLA}
S.~L.~Glashow, 
   Nucl.~Phys.~{\bf 22}, 579 (1961).
\bibitem{S1}
A.~Salam, 
   in {\it Proc.~8th Nobel Symposium on Weak and Electromagnetic 
           Interaction} (Svartholm, 1968), p.~367.
\bibitem{W2}
S.~Weinberg, 
   Phys.~Rev.~Lett.~{\bf 19}, 1264 (1967).
\bibitem{SB1}
J.~Goldstone, 
   Nuovo~Cim.~{\bf 19}, 154 (1961).
\bibitem{SB2}
P.~W.~Higgs,  
   Phys.~Rev.~Lett.~{\bf 13}, 508 (1964); 
   Phys.~Rev.~{\bf 145}, 1156 (1966). 
\bibitem{swm}
S.~De Leo,~Z.~Oziewicz,~W.~A.~Rodrigues and J.~Vaz,  
   {\em Space-Time Algebra and Salam-Weinberg Model} 
   (in preparation).

\end{references}
\end{document}